\documentclass[twoside,fleqn]{ActaStyle}
\usepackage{times,cite}
\usepackage{graphicx,epsfig}    

\def\authorheading{M. Kaushik, D. Singh and H.L. Yadav }
\def\shorttitle{Halo formation in neutron rich Ca Isotopes}
\begin{document}
\pagerange{1}{16}
\title{HALO FORMATION IN NEUTRON RICH $\rm{Ca}$ NUCLEI}
\author{  M. Kaushik, D. Singh, H. L. Yadav
\email{hlyadav@sancharnet.in}} { Department of Physics, Rajasthan
University, Jaipur 302004, India}
%
\begin{center}
Received 22 April 2004 accepted 6 December 2004
\end{center}
\abstract{ We have investigated the halo formation in the neutron
rich $\rm{Ca}$ isotopes within the framework of recently proposed
relativistic mean-field plus BCS (RMF+BCS) approach wherein the
single particle continuum corresponding to the RMF is replaced by
a set of discrete positive energy states for the calculation of
pairing energy. For the neutron rich $\rm{Ca}$ isotopes in the
vicinity of neutron drip-line, it is found that further addition
of neutrons causes a rapid increase in the neutron rms radius with
a very small increase in the binding energy, indicating thereby
the occurrence of halos. This is essentially caused by the gradual
filling in of the loosely bound $3s_{1/2}$ state. Interesting
phenomenon of accommodating several additional neutrons with
almost negligible increase in binding energy is shown to be due to
the pairing correlations.}

\vspace*{2pt}

\section{Introduction}
\label{sec:intr} \setcounter{section}{1}\setcounter{equation}{0}

\noindent  The availability of radioactive beam facilities has
generated a spurt of activity devoted to the investigation of exotic
drip line nuclei. The  neutron rich nuclei away from the line of
$\beta$-stability with unusually large isospin value are known to
exhibit several interesting features.  For nuclei close to the
neutron drip-line, the neutron density distribution shows a much
extended tail with  a diffused neutron skin while the Fermi level
lies close to the single particle continuum \cite{tanihata}. In some
cases it may even lead to the phenomenon of neutron halo, as
observed in the case of light nuclei
\cite{tanihata,jensen,ring0,dobac0} , made of several neutrons
outside a core with separation energy of the order of $\approx$ 100
keV or less. Interestingly, a theoretical discussion on the
possibility of occurrence of such structures has been considered by
Migdal\cite{migdal73} already in early 70's. Obviously for such
nuclei, due to the weak binding and large spatial dimension of the
outermost nucleons, the role of continuum states and their coupling
to the bound states become exceedingly important, especially for the
pairing energy contribution to the total binding energy of the
system. Theoretical investigations of such neutron rich nuclei have
been carried out extensively within the framework of  mean field
theories \cite{dobac,terasaki,dobac96,sand,sand2,grasso} and also
employing their relativistic
counterparts\cite{ring0,walecka,serot,pgr,bouy,pgr2,gambhir,toki,
brock,hirata,suga,ring,sharma,meng,meng4,lala,mizu,
leja,estal,yadav,yadav1,meng2}.

Recently the effect of continuum on the pairing energy
contribution has been studied by Grasso {\it et al}.\cite{grasso}
and Sandulescu {\it et al}.\cite{sand2} within the HF+BCS+Resonant
continuum approach. Similarly the effect of inclusion of positive
energy resonant states on the pairing correlations has been
investigated by Yadav et al.\cite{yadav}. A detailed comparative
study of the Hartree-Fock-Bogoliubov (HFB) approach with those of
the HF+BCS+Resonant continuum calculations carried out by Grasso
{\it et al}.\cite{grasso} and Sandulescu {\it et al}.\cite{sand2}
has provided useful insight to the validity of different
approaches for the treatment of drip-line nuclei. The interesting
result of these investigation is that only a few low energy
resonant states, especially those near the Fermi surface influence
in an appreciable way  the pairing properties of nuclei far from
the $\beta$-stability.
This finding is of immense significance because one can eventually
make use of this for systematic studies of a large number of nuclei
by employing a simpler HF+BCS approximation.  Amongst the mean-field
theoretic treatments, however, currently the relativistic mean field
(RMF) theory is being extensively used for the study of unstable
nuclei \cite{ring0,toki,brock,hirata,suga,ring,sharma,meng,meng4}. The
advantage of the RMF approach is that it provides the spin-orbit
interaction in the entire mass region in a natural way
\cite{walecka,serot,pgr}. This indeed has proved to be very crucial
for the study of unstable nuclei near the drip line, since the
single particle properties near the threshold are prone to large
changes as compared to the case of deeply  bound levels in the
nuclear potential. In addition to this, the pairing properties are
equally important for nuclei near the drip line. In order to take
into account the pairing correlations together with a realistic mean
field, the framework of standard RHB approach is commonly
used\cite{meng,lala}. In this connection, the finding above for the
non-relativistic frameworks has turned out to be very important for
the systematic work of unstable nuclei in the relativistic approach.
This has been demonstrated recently by Yadav et
al.\cite{yadav,yadav1} for the chains of $^{48-98}\rm{Ni}$ and
$^{96-176}\rm{Sn}$ isotopes covering the drip lines. Indeed the
RMF+BCS scheme\cite{yadav,yadav1} wherein the single particle
continuum corresponding to the RMF is replaced by a set of discrete
positive energy states yields results which are found to be in close
agreement with the experimental data and with those of recent
continuum relativistic Hartree-Bogoliubov (RCHB) and other similar
mean-field calculations\cite{meng,meng2}.

With the success of the RMF+BCS approach for the prototype
calculations of $\rm{Ni}$ and $\rm{Sn}$ isotopes\cite{yadav,
yadav1}, detailed calculations for the chain of $\rm{Ca}$ isotopes
and also those of $\rm{O,\, Ni,\, Zr,\, Sn }$ and $\rm{Pb}$
isotopes using the TMA\cite{suga} and the NL-SH\cite{sharma} force
parameterizations have been carried out. The results of these
calculations\cite{yadav2} for the two neutron separation energy,
neutron, proton, and matter rms radii, and single particle pairing
gaps etc., and their comparison with the available experimental
data and with the results of other mean-field approaches
demonstrate the general validity of the RMF+BCS approach. In this
paper, in continuation to our earlier publication\cite{yadav1}, we
present briefly the results  for the chain of $\rm{Ca}$ isotopes
but with a special emphasis on our findings for the possible halo
formation in the neutron rich $\rm{Ca}$ isotopes within the
RMF+BCS approach. It is shown that the resonant $1g_{9/2}$ and the
$3s_{1/2}$ states which lie close to zero energy in continuum and
gradually come down to become bound with increasing neutron
number, play the crucial role.
Evidently the concentration of the major part of the wave function
of the resonant $1g_{9/2}$ state within the potential well and its
proximity with the Fermi surface while being close to zero energy
together provide a favorable condition for the existence of
extremely neutron rich $\rm{Ca}$ isotopes, whereas the $3s_{1/2}$
state with a well spread wave function, due to the absence of a
centrifugal barrier, helps to cause the occurrence of halos. The
role of pairing correlations as described here is found to be
consistent with the conclusions of non-relativistic HFB studies of
neutron rich weakly bound nuclei discussed recently by Bennaceur
et al. \cite{dobac0}.

\section{Theoretical Formulation and Model}

\noindent Our RMF calculations have been carried out using the
model Lagrangian density with nonlinear terms both for the
${\sigma}$ and ${\omega}$ mesons as described in detail in
Refs.~\cite{suga}.

\begin{eqnarray}
       {\cal L}& = &{\bar\psi} [\imath \gamma^{\mu}\partial_{\mu}
                  - M]\psi\nonumber\\
                  &&+ \frac{1}{2}\, \partial_{\mu}\sigma\partial^{\mu}\sigma
                - \frac{1}{2}m_{\sigma}^{2}\sigma^2- \frac{1}{3}g_{2}\sigma
                  ^{3} - \frac{1}{4}g_{3}\sigma^{4} -g_{\sigma}
                 {\bar\psi}  \sigma  \psi\nonumber\\
                &&-\frac{1}{4}H_{\mu \nu}H^{\mu \nu} + \frac{1}{2}m_{\omega}
                   ^{2}\omega_{\mu}\omega^{\mu} + \frac{1}{4} c_{3}
                  (\omega_{\mu} \omega^{\mu})^{2}
                   - g_{\omega}{\bar\psi} \gamma^{\mu}\psi
                  \omega_{\mu}\nonumber\\
               &&-\frac{1}{4}G_{\mu \nu}^{a}G^{a\mu \nu}
                  + \frac{1}{2}m_{\rho}
                   ^{2}\rho_{\mu}^{a}\rho^{a\mu}
                   - g_{\rho}{\bar\psi} \gamma_{\mu}\tau^{a}\psi
                  \rho^{\mu a}\nonumber\nonumber\\
                &&-\frac{1}{4}F_{\mu \nu}F^{\mu \nu}
                  - e{\bar\psi} \gamma_{\mu} \frac{(1-\tau_{3})}
                  {2} A^{\mu} \psi\,\,,
\end{eqnarray}
where the field tensors $H$, $G$ and $F$ for the vector fields are
defined by
\begin{eqnarray}
                 H_{\mu \nu} &=& \partial_{\mu} \omega_{\nu} -
                       \partial_{\nu} \omega_{\mu}\nonumber\\
                 G_{\mu \nu}^{a} &=& \partial_{\mu} \rho_{\nu}^{a} -
                       \partial_{\nu} \rho_{\mu}^{a}
                     -2 g_{\rho}\,\epsilon^{abc} \rho_{\mu}^{b}
                    \rho_{\nu}^{c} \nonumber\\
                  F_{\mu \nu} &=& \partial_{\mu} A_{\nu} -
                       \partial_{\nu} A_{\mu}\,\,,\nonumber\
\end{eqnarray}
and other symbols have their usual meaning.

The set of parameters appearing in the effective Lagrangian (1)
have been obtained in an extensive study which provides a good
description for the ground state of nuclei and that of the nuclear
matter properties\cite{suga}. This set, termed as TMA, has an
$A$-dependence and covers the light as well as heavy nuclei from
$^{16}\rm{O}$ to $^{208}\rm{Pb}$. Table 1 lists the TMA set of
parameters along with the results for the calculated bulk
properties of nuclear matter. As mentioned earlier we have also
carried out the RMF+BCS calculations using the NL-SH force
parameters \cite{sharma}  in order to compare our results with
those obtained in the RCHB calculations \cite{meng2} using this
force parameterizations. The NL-SH parameters are also listed in
Table 1 together with the corresponding nuclear matter properties.

Based on the single-particle spectrum calculated by the RMF
described above, we perform a state dependent BCS
calculations\cite{lane,ring2}. As we already mentioned, the
continuum is replaced by a set of positive energy states generated
by enclosing the nucleus in a spherical box. Thus the gap
equations have the standard form for all the single particle
states, i.e.
\begin{eqnarray}
     \Delta_{j_1}& =&\,-\frac{1}{2}\frac{1}{\sqrt{2j_1+1}}
     \sum_{j_2}\frac{\left<{({j_1}^2)\,0^+\,|V|\,({j_2}^2)\,0^+}\right>}
      {\sqrt{\big(\varepsilon_{j_2}\,-\,\lambda \big)
       ^2\,+\,{\Delta_{j_2}^2}}}\,\,\sqrt{2j_2+1}\,\,\, \Delta_{j_2}\,\,,
\end{eqnarray}\\
where $\varepsilon_{j_2}$ are the single particle energies, and
$\lambda$ is the Fermi energy, whereas the particle number
condition is given by $\sum_j \, (2j+1) v^2_{j}\,=\,{\rm N}$. In
the calculations we use for the pairing interaction a delta force,
i.e., $V=-V_0 \delta(r)$ with the same strength $V_0$ for both
protons and neutrons. The value of the interaction strength $V_0 =
350\,$ MeV fm$^3$ was determined in ref. \cite{yadav} by obtaining
a best fit to the  binding energy of $\rm{Ni}$ isotopes. We use
the same value of $V_0$ for our present studies of isotopes
\begin{table}[htb]
\centering \caption{Parameters of the Lagrangian TMA\cite{suga}
and NL-SH\cite{sharma} together with the nuclear matter properties
obtained with these effective forces.}
\bigskip
\begin{tabular}
{|lcc|ccc|}
\hline \multicolumn{1}{|c}{}& \multicolumn {1}{c}{Force
Parameters}&\multicolumn {1}{c|}{}&\multicolumn{2}{c}
{}{Nuclear Matter Properties}& \multicolumn{1}{c|}{}\\
\cline{1-6}
{}&{TMA}&{NL-SH}&{}&{TMA}&{NL-SH}\\
\cline{2-6} \hline
M (MeV)&938.9&939.0&Saturation density&&   \\
m$_{\sigma}$(MeV) &519.151&526.059&$\rho_{0}$ (fm)$^{-3}$&0.147&0.146\\
\cline{4-6}
m$_{\omega}$(MeV) &781.950&783.0&Bulk binding energy/nucleon  &&\\
m$_{\rho}$(MeV) &768.100&763.0&(E/A)$_{\infty}$ (MeV)&16.0&16.346\\
\cline{4-6}
g$_{\sigma}$&  10.055 + 3.050/A$^{0.4}$&10.444&Incompressibility&& \\
g$_{\omega}$&  12.842 + 3.191/A$^{0.4}$&12.945&K (MeV)&318.0& 355.36 \\
\cline{4-6}
g$_{\rho}$ &   3.800 + 4.644/A$^{0.4}$&4.383&Bulk symmetry energy/nucleon&& \\
g$_{2}$ (fm)$^{-1}$ & -0.328 - 27.879/A$^{0.4}$&-6.9099& a$_{sym}$ (MeV)&30.68& 36.10\\
\cline{4-6}
g$_{3}$ & 38.862 - 184.191/A$^{0.4}$&-15.8337&Effective mass ratio &&\\
c$_{3}$ &151.590 - 378.004/A$^{0.4}$&& m$^{*}$/m &0.635& 0.60  \\
\hline
\end{tabular}
\end{table}
of other nuclei as well. Apart from its simplicity, the applicability
and justification of using such a $\delta$-function form of
interaction has been recently discussed in Refs.\cite{dobac} and
\cite{dobac96}, whereby it has been shown in the context of HFB
calculations that the use of a delta force in a finite space
simulates the effect of finite range interaction in a
phenomenological manner ( see also \cite{bertsch91} and
\cite{migdal67} for more details ). The pairing matrix element for
the $\delta$-function force is given by\

\begin{eqnarray}
\left<{({j_1}^2)\,0^+\,|V|\,({j_2}^2)\,0^+}\right>&
=&\,-\,\frac{V_0}{8\pi}
       \sqrt{(2j_1+1)(2j_2+1)}\,\,I_R\,\,,
\end{eqnarray}
where $I_R$ is the radial integral having the form
\begin{eqnarray}
   I_R& =&\,\int\,dr \frac{1}{r^2}\,\left(G^\star_{j_ 1}\, G_{j_2}\,+\,
     F^\star_{j_ 1}\, F_{j_2}\right)^2
\end{eqnarray}
Here $G_{\alpha}$ and $F_{\alpha}$ denote the radial wave
functions for the upper and lower components, respectively, of the
nucleon wave function expressed as
\begin{equation}\psi_\alpha={1 \over r} \,\, \left({i \,\,\, G_\alpha \,\,\,
 {\mathcal Y}_{j_\alpha l_\alpha m_\alpha}
\atop{F_\alpha \, {\sigma} \cdot \hat{r}\, \, {\mathcal
Y}_{j_\alpha l_\alpha m_\alpha}}} \right)\,\,,
\end{equation}
and satisfy the normalization condition
 \begin{eqnarray}
         \int dr\, {\{|G_{\alpha}|^2\,+\,|F_{\alpha}|^2}\}\,=\,1
 \end{eqnarray}\
In Eq. (5) the symbol ${\mathcal Y}_{jlm}$ has been used for the
standard spinor spherical harmonics with the phase $i^l$. The
coupled field equations obtained from the Lagrangian density in
(1) are finally reduced to a set of simple radial
equations\cite{pgr} which are solved self consistently along with
the equations  for the state dependent pairing gap $\Delta_{j}$
and the total particle number $\rm N$ for a given nucleus.

\begin{center}
\vskip 0.01in
\psfig{figure=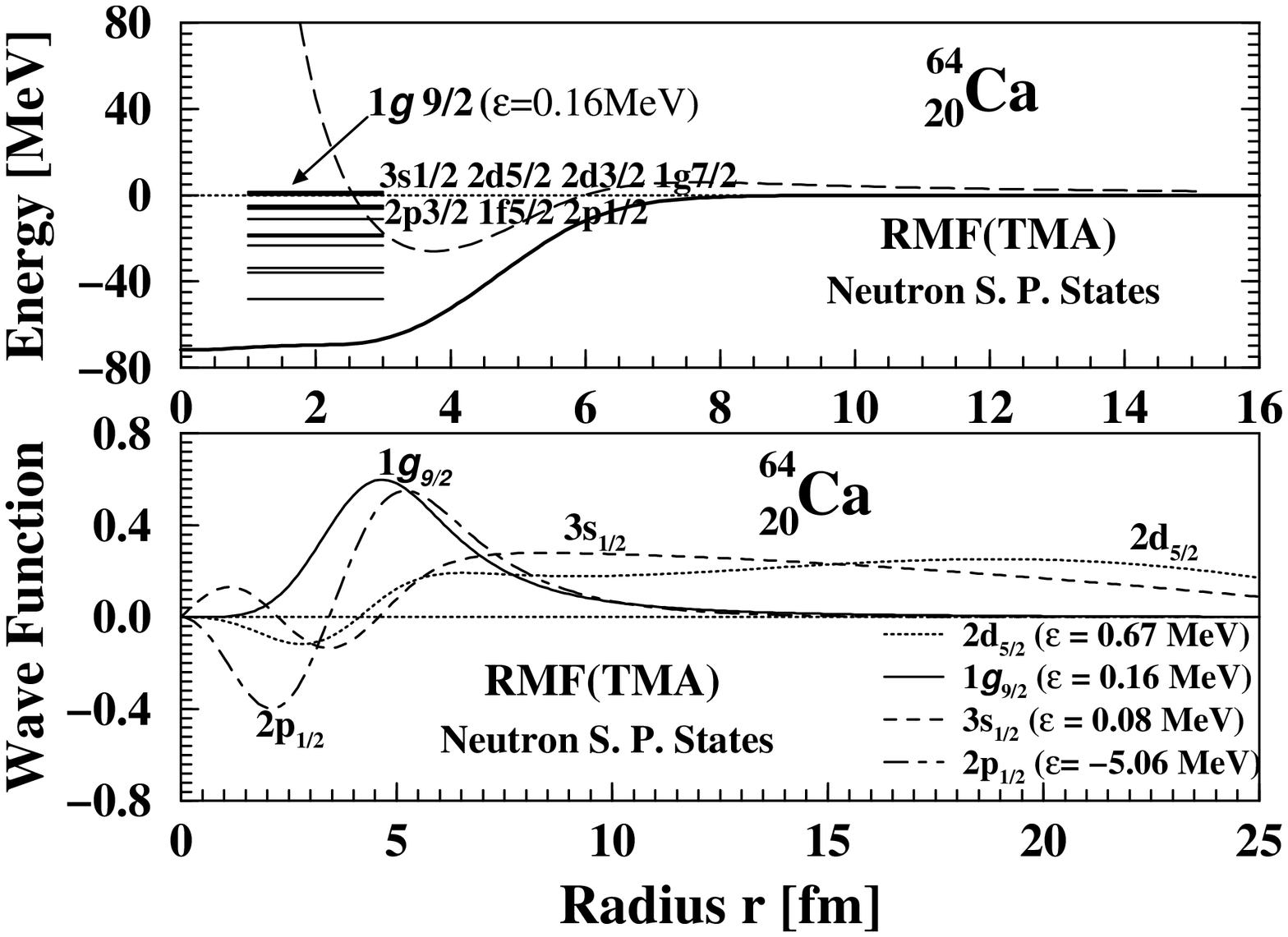,width=10.0cm,height=08.0cm}
\vskip 0.01in {\noindent \small {{\bf Fig. 1.} Upper panel: The
RMF potential energy (sum of the scalar and vector potentials),
for the nucleus $^{64}\rm{Ca}$ shown by the solid line as a
function of radius $r$. The long dashed line represents the sum of
RMF potential energy and the centrifugal barrier energy for the
neutron resonant state $1g_{9/2}$. It also shows the energy
spectrum of some important neutron single particle states along
with the resonant $1g_{9/2}$ state at 0.16 MeV. Lower panel:
Radial wave functions of a few representative neutron single
particle states with energy close to the Fermi surface for the
nucleus $^{64}\rm{Ca}$. The solid line shows the resonant
$1g_{9/2}$ state. \mbox{} }}
\end{center}
\section{Results and Discussion}
Our earlier calculations for chains of $\rm{Ni}$ and $\rm{Sn}$
isotopes\cite{yadav,yadav1} and  present investigations of
$\rm{Ca}$ isotopes as well as of other nuclei indicate that the
neutron rich $\rm{Ca}$ isotopes constitute the most interesting
example of loosely bound system. For an understanding of such an
exotic system the total pairing energy contribution to the binding
energy plays a  crucial role. This in turn implies the importance
of the structure of single particle states near the Fermi level,
as the scattering of particles from bound to continuum states and
vice versa due to pairing interaction involves mainly these
states. The last few occupied states near the Fermi level also
provide an understanding of the radii of the loosely bound exotic
nuclei. The neutron rich nuclei in which the last filled single
particle state near the Fermi level is of low angular momentum
($s_{1/2}$ or $p_{1/2}$ state), especially the $l=0$ state, can
have large radii due to large spatial extension of the $s_{1/2}$
state which has no centrifugal barrier.

In order to demonstrate our results we have chosen $^{64}\rm{Ca}$
as a representative example of the neutron rich $\rm{Ca}$
isotopes. Moreover, since the results obtained with TMA and NL-SH
forces are found to be almost similar, to save space we describe
in details only the results for the TMA force, whereas the results
obtained with the NL-SH force have been discussed at places for
the purpose of comparison. The upper panel of Fig.1 shows the
calculated RMF potential, a sum of scalar and vector potentials,
along with the spectrum for the bound neutron single particle
states for the neutron rich $^{64}\rm{Ca}$ obtained with the TMA
force. The figure also shows the positive energy state
corresponding to the first low-lying resonance $1g_{9/2}$, and
other positive energy states, for example, $3s_{1/2}$, $2d_{5/2}$,
$2d_{3/2}$ and $1g_{7/2}$ close to the Fermi surface which play
significant role for the binding of neutron rich isotopes through
their contributions to the total pairing energy. In contrast to
other states in the box which correspond to the non-resonant
continuum, the position of the resonant $1g_{9/2}$ state is not
much affected by changing the box radius around $R=30$ fm. We have
also depicted in this part of Fig.1 the total mean field potential
for the neutron $1g_{9/2}$ state, obtained by adding the
centrifugal potential energy. It is evident from the figure that
the effective total potential for the $1g_{9/2}$ state has an
appreciable barrier to form a quasi-bound or resonant state. Such
a meta-stable state remains mainly confined to the region of the
potential well and the wave function exhibits characteristics
similar to that of a bound state. This is clearly seen in the
lower panel of Fig.1 which depicts the radial wave functions of
some of the neutron single particle states lying close to the
Fermi surface, the neutron Fermi energy being $\lambda_n\,
=\,-0.066$ MeV. These include the bound  $2p_{1/2}$, and the
continuum $3s_{1/2}$ and $2d_{5/2}$  states in addition to the
state corresponding to the resonant $1g_{9/2}$. The wave function
for the $1g_{9/2}$ state in Fig. 1 (lower panel) is clearly seen
to be confined within a radial range of about 8 fm and has a
decaying component outside this region, characterizing a resonant
state. In contrast, the main part of the wave function for the
non-resonant states, e.g. $2d_{5/2}$, is seen to be mostly spread
over outside the potential region. This type of state thus has a
poorer overlap with the bound states near the Fermi surface leading
to small value for the pairing gap $\Delta_{2d_{5/2}}$. Further,
the positive energy states lying much higher from the Fermi level, for
example, $1h_{11/2}$, $1i_{13/2}$ etc.  have a negligible contribution to
the total pairing energy of the system.These features can be seen from
Fig.2 (upper panel) which depicts pairing gap energy $\Delta_j$ for the
neutron states in $^{64}\rm{Ca}$. The gap energy for the $1g_{9/2}$ state
is seen to have a value close to $1$ MeV which is quantitatively similar to
that of bound states $1f_{7/2}$ and $2p_{3/2}$ etc. The
non-resonant states like $3s_{1/2}$ and $2d_{5/2}$ in continuum
have much smaller gap energy. However, while approaching the
neutron drip line nucleus $^{72}\rm{Ca}$, the single particle
states $3s_{1/2}$, $1g_{9/2}$, $2d_{5/2}$ and $2d_{3/2}$ which lie
near the Fermi level gradually come down close to zero energy, and
subsequently the $1g_{9/2}$ and $3s_{1/2}$ states become bound.
This helps in accommodating more and more neutrons with very
little binding. In fact, the occupancy of the $3s_{1/2}$ state in
these neutron rich isotopes causes the halo formation as will be
seen  later.
In the lower panel of Fig.2 we have shown the contribution of
pairing energy which plays an important role for the stability of
nuclei and consequently in deciding the position of the neutron
and proton drip lines. It is seen that the RMF+BCS calculations
carried out with two different sets of force parameters, the TMA
and NL-SH, yield almost similar results also for the pairing
energies. The differences in the two results can be attributed to
the difference in the detailed structure of single particle
energies obtained with the TMA and NL-SH forces. One observes from
Fig.2 (lower panel) that the pairing energy vanishes for the
neutron numbers $N = 14, 20, 28$ and $40$ indicating the shell
closures. In particular the usual shell closure at $N=50$ is found
to be absent for the neutron rich $\rm{Ca}$ isotopes and at $N=40$
a new shell closure appears. This reorganization of single
particle energies with large values of N/Z ratio (for the neutron
rich $\rm{Ca}$ isotopes N/Z $\geq 2$) has its origin in the
deviation of the strength of spin-orbit splitting from the
conventional shell model results for nuclei with not so large
$N/Z$ ratio.

\begin{center}
\vskip 0.01in \psfig{figure=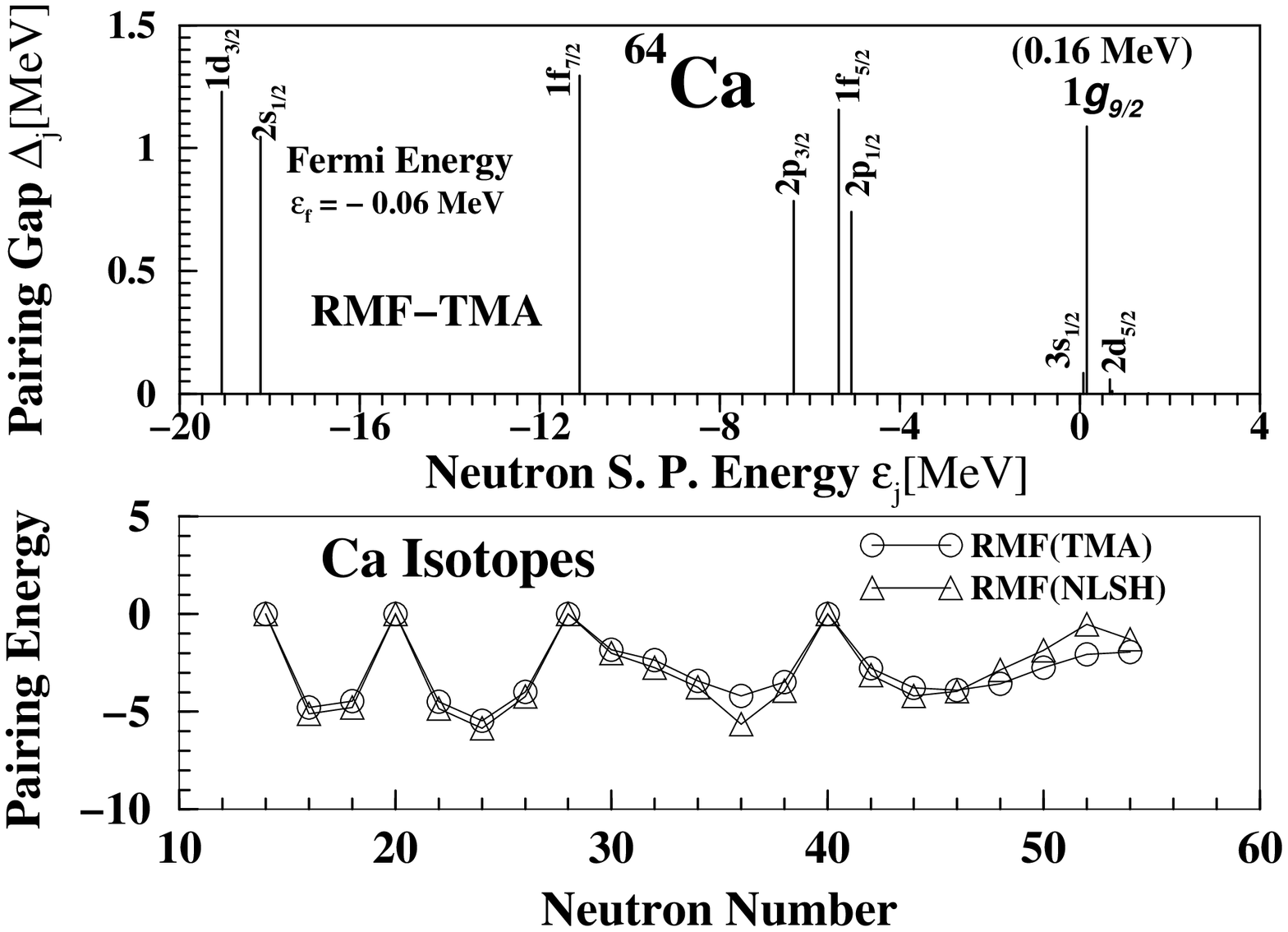,width=10.0cm,height=08.0cm}
\vskip 0.01in {\noindent \small {{\bf Fig.2.} Upper panel: Pairing
gap energy $\Delta_{j}$ of neutron single particle states with
energy close to the Fermi surface for the nucleus $^{64}\rm{Ca}$.
The resonant $1g_{9/2}$ state at energy 0.165 MeV has the gap
energy of about $1$ MeV which is close to that of bound states
like $1f_{5/2}$, $1f_{7/2}$, $1d_{3/2}$ etc.\ Lower panel: Pairing
energy for the $\rm{Ca}$ isotopes obtained with the TMA (open
circles) and the NL-SH (open triangles) force parameters. \mbox{}
}}
\end{center}

The results for the ground state properties including the binding
energy, two neutron separation energy, and the rms radii for the
neutron, proton, matter and charge distributions for the $\rm{Ca}$
isotopes calculated with the TMA force have been listed in Table
2. The table also lists the available experimental data\cite{audi}
for the binding energies, two neutron separation energy as well as
the binding energy per nucleon (B/A) for the purpose of
comparison. The available experimental data for the radii are very
sparse and have not been listed. It is interesting to note that
the maximum value of B/A occurs for the $^{46}\rm{Ca}$ isotope and
not the doubly magic isotopes $^{40}\rm{Ca}$ or $^{48}\rm{Ca}$.
The shell structure as revealed by the pairing energies are also
exhibited in the variation of two neutron separation energies
$S_{2n}$ as shown in the lower panel of Fig.3. An abrupt increase
in the $S_{2n}$ values for the isotopes next to magic numbers is
clearly seen. This part of the figure also depicts the results of
two neutron separation energy obtained in the RCHB approach and
their comparison  with available experimental data. The upper
panel of Fig.3 depicts the difference between the experimental and
calculated values. It is seen that the TMA and NL-SH forces yield
similar results and the isotopes beyond A$=$66 have two neutron
separation energy close to zero. The two neutron drip line is
found to occur at A$=$70 (N$=$50) and A$=$72 (N$=$52) for the TMA
and  NL-SH forces, respectively.  The isotopes with mass number
$70<A<76$ for the TMA, and those with $72<A<76$ for the NL-SH case
are found to be just unbound with negative separation energy very
close to zero. Accordingly in Fig.3 we have shown the results up
to N$=$56 to emphasize this point. Also, for our purpose we shall
neglect this small difference in the position of drip line
mentioned above. The similarity of different calculated results
amongst themselves and their satisfactory comparison with data are
well demonstrated in the upper panel.

\begin{table}[htb]
\centering \caption{Results for the ground state properties of
Ca-isotopes calculated with the TMA force parameter set. Listed
are the total binding energy, BE, the two neutron separation
energy, S$_{2n}$, binding energy per nucleon, B/A, and neutron,
proton, matter and charge root mean square radii denoted by r$_n$,
r$_p$, r$_m$, r$_c$, respectively. The available experimental
data\cite{audi} on the binding energy, (BE)$_{exp}$, and that of
(S$_{2n}$)$_{exp}$ and (B/A)$_{exp}$ are also listed for
comparison.}
\bigskip
\begin{tabular}{|c|c|c|c|c|c|c|c|c|c|c|}
 \hline \multicolumn{1}{|c|}{Nucleus}&
\multicolumn{1}{|c|}{(BE)$_{exp}$}& \multicolumn{1}{|c|}{BE}&
\multicolumn{1}{|c|}{(S$_{2n}$)$_{exp}$}&
\multicolumn{1}{|c|}{S$_{2n}$}&\multicolumn{1}{|c|}{(B/A)$_{exp}$}&
\multicolumn{1}{|c|}{B/A}&\multicolumn{1}{|c|}{r${_n}$}&
\multicolumn{1}{|c|}{r${_p}$}&
\multicolumn{1}{|c|}{r${_m}$}& \multicolumn{1}{|c|}{r${_c}$}\\
\multicolumn{1}{|c|}{}& \multicolumn{1}{|c|}{MeV}&
\multicolumn{1}{|c|}{MeV}& \multicolumn{1}{|c|}{MeV}&
\multicolumn{1}{|c|}{MeV}&\multicolumn{1}{|c|}{MeV}&
\multicolumn{1}{|c|}{MeV}&\multicolumn{1}{|c|}{fm}&
\multicolumn{1}{|c|}{fm}&
\multicolumn{1}{|c|}{fm}& \multicolumn{1}{|c|}{fm}\\
\hline
$^{34}Ca$&245.625&246.684&{}&{}&7.224&7.255&3.039&3.415&3.266&3.511 \\
$^{36}Ca$&281.360&280.968&35.735&34.284&7.816&7.805&3.168&3.394&3.296&3.489 \\
$^{38}Ca$&313.122&313.114&31.762&32.146&8.240&8.240&3.262&3.385&3.327&3.479 \\
$^{40}Ca$&342.052&343.208&28.930&30.094&8.551&8.580&3.337&3.383&3.360&3.475 \\
$^{42}Ca$&361.895&363.843&19.843&20.635&8.617&8.663&3.426&3.381&3.404&3.471 \\
$^{44}Ca$&380.960&382.823&19.065&18.980&8.658&8.700&3.501&3.382&3.447&3.471 \\
$^{46}Ca$&398.769&400.385&17.809&17.562&8.669&8.704&3.565&3.386&3.488&3.473 \\
$^{48}Ca$&415.991&416.629&17.222&16.244&8.666&8.680&3.621&3.391&3.527&3.476 \\
$^{50}Ca$&427.491&425.235&11.500&8.606&8.550&8.505&3.759&3.412&3.624&3.495 \\
$^{52}Ca$&436.600&433.287&9.109&8.052&8.396&8.332&3.872&3.435&3.710&3.516 \\
$^{54}Ca$&443.800&440.711&7.200&7.424&8.219&8.161&3.966&3.463&3.788&3.542 \\
$^{56}Ca$&449.600&448.148&5.800&7.437&8.029&8.003&4.082&3.493&3.882&3.569 \\
$^{58}Ca$&&455.380&&7.232&&7.851&4.115&3.522&3.921&3.596 \\
$^{60}Ca$&&462.135&&6.755&&7.702&4.173&3.551&3.977&3.623 \\
$^{62}Ca$&&462.704&&0.569&&7.463&4.248&3.573&4.042&3.643\\
$^{64}Ca$&&462.964&&0.260&&7.234&4.361&3.593&4.137&3.662 \\
$^{66}Ca$&&463.032&&0.068&&7.016&4.634&3.609&4.349&3.675 \\
$^{68}Ca$&&463.058&&0.026&&6.810&4.851&3.623&4.525&3.687 \\
$^{70}Ca$&&463.075&&0.017&&6.615&4.926&3.640&4.596&3.703 \\
$^{72}Ca$&&462.972&&-0.103&&6.430&5.007&3.655&4.671&3.716 \\
\hline
\end{tabular}
\end{table}
\vskip 0.075in
The rms radii for the proton and neutron , $r_{p,n}\, =\,(\langle
r^2_{p(n)}\rangle\,)^{1/2} $ have been calculated from the
respective density distributions. The experimental data for the
rms charge radii are used to deduce the nuclear rms proton radii
using the relation $r_c^2\,=\,r_p^2\,+\,0.64 \, fm^2$ for the
purpose of comparison. In the middle panel of Fig.4 we have shown
the RMF+BCS results for the neutron and proton rms radii for the
NL-SH force along with the RCHB results\cite{meng2} also obtained
using the NL-SH force for the purpose of comparison. These results
are quite similar as can be seen from the differences plotted in
the upper panel. The experimental data for the proton and neutron
rms radii are available only for a few stable $\rm Ca$ isotopes.
The lower panel of Fig.4 depicts a comparison of the RMF+BCS
results using the TMA force with the available experimental data.
It is seen from Fig.4 that the measured proton radii $r_p$ for the
isotopes $^{40-48}\rm Ca$ are in excellent agreement with our
RMF+BCS results. Similarly the neutron radii $r_n$ for the
$^{40,42,44,48}\rm Ca$ isotopes are found to compare quite well as
has been depicted in the lower panel of Fig.4.

As described earlier, in the case of neutron rich $\rm Ca$
isotopes the neutron $1g_{9/2}$ state happens to be a resonant
state having good overlap with the bound states near the Fermi
level. This causes the pairing interaction to scatter particles
from the neighboring bound states to the resonant state

\psfig{figure=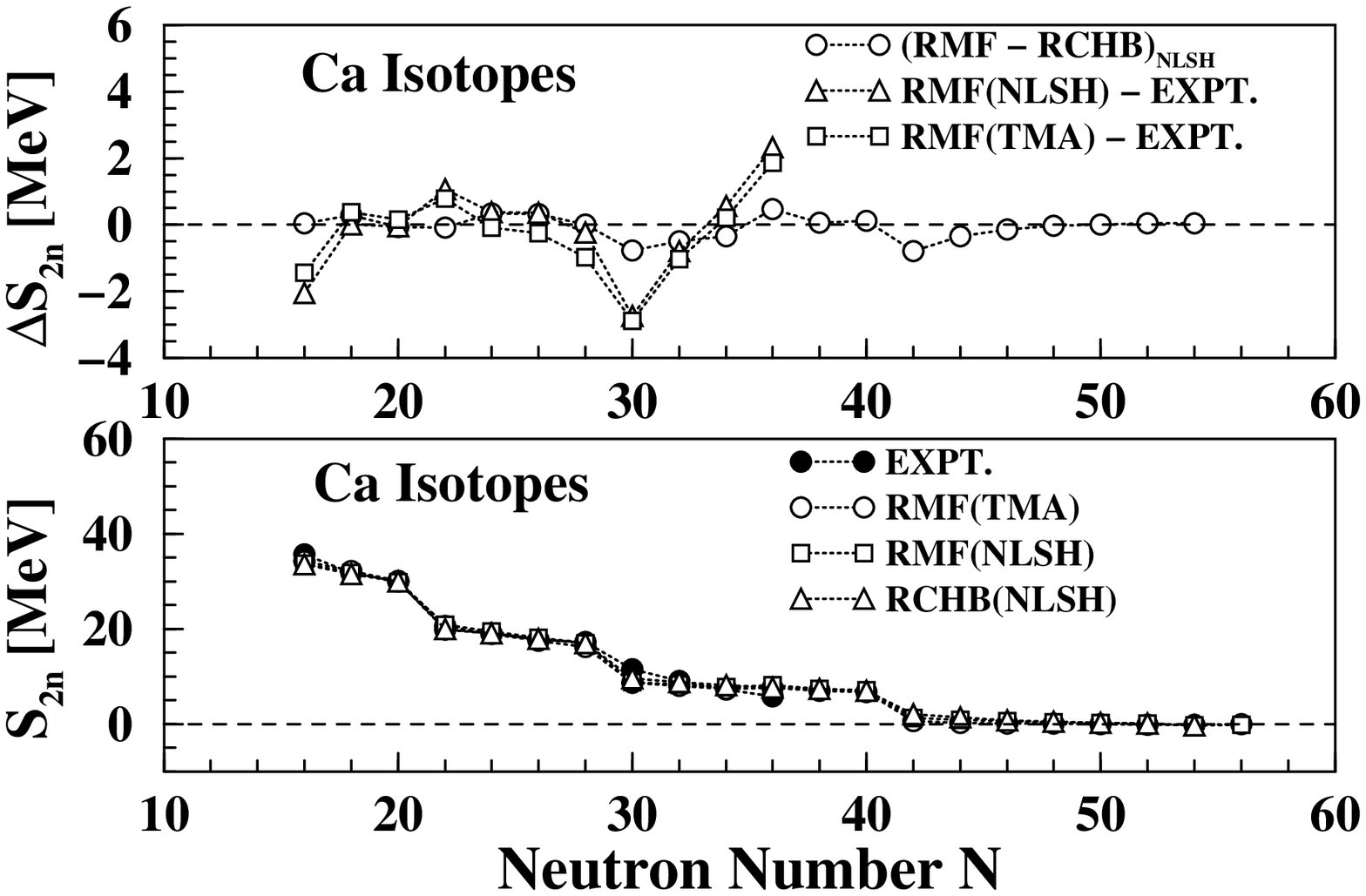,width=10.0cm,height=08.0cm}
\vskip0.001in {\noindent \small {{\bf Fig.3.} In the lower panel two
neutron separation energies for the $\rm{Ca}$ isotopes calculated
with the TMA (open circles) and the NL-SH (open squares) force
parameters are compared with the RCHB calculations of
Ref.\cite{meng2} carried out with the NL-SH force (open triangles)
and with the available experimental data\cite{audi}. The upper
panel shows the difference in the RMF+BCS and the RCHB results as
well as  the difference between calculated results with respect to
the available experimental data\cite{audi}.\mbox{} }}
\vskip0.11in
\noindent and vice versa. Thus, it is found that the resonant $1g_{9/2}$ state starts
being partially occupied even before the lower bound single
particle states are fully filled in. Further, for neutron rich
$\rm Ca$ isotopes it is found that the neutron $3s_{1/2}$ state
which lies close to the $1g_{9/2}$ state also starts getting
partially occupied before the $1g_{9/2}$ state is completely
filled. The neutron $3s_{1/2}$ state due to lack of centrifugal
barrier contributes more to the neutron rms radius as compared to
the $1g_{9/2}$ state, and thus one observes in Fig.4 (lower panel)
a rapid increase in the neutron rms radius beyond the neutron
number $N=42$ indicating the formation of halos. A comparison of
the rms neutron radii with the $r_n=r_0 N^{1/3}$ line shown in the
figure suggests that these radii for the drip-line isotopes do not
follow the $r_0 N^{1/3}$ systematics.

Similar calculations for the neutron radii of $\rm Ni$ and $\rm
Sn$ isotopes\cite{yadav,yadav1}, however, do not exhibit halo
formation as can be seen from Fig.5. As regards the possibility of
halo formation in other nuclei, the following remark is pertinent.
From our comprehensive calculations of chains of proton magic
nuclei\cite{yadav2}, it is found that the isotopes of $\rm O$ as
well as $\rm Pb$ nuclei also do not exhibit the tendency of halo
formations. On the other hand, the proton sub-magic neutron rich
$\rm Zr$ isotopes do have a single particle structure that
provides a favorable condition for the halos, albeit in a less
pronounced manner. In this case the single particle $3p_{1/2}$
state lying close to the continuum threshold plays the crucial
role.

An important aspect of the heavy neutron rich nuclei is the
formation of the neutron skin.\cite{tanihata} The neutron density
distributions in the neutron rich $^{62-72}\rm{Ca}$ nuclei are
found to be widely spread out in the space indicating the
formation of neutron halos. This has been demonstrated in Fig.6
which shows the variation in the proton (upper panel) and neutron
(lower panel) radial density distributions with increasing neutron
number for the TMA force calculations. It may be emphasized that
the density distributions obtained from the NL-SH force
parameterizations are almost similar to those obtained using the
TMA force and, therefore, here we have chosen to show only the
results for the TMA force.
\begin{center}
\vskip 0.0015in \psfig{figure=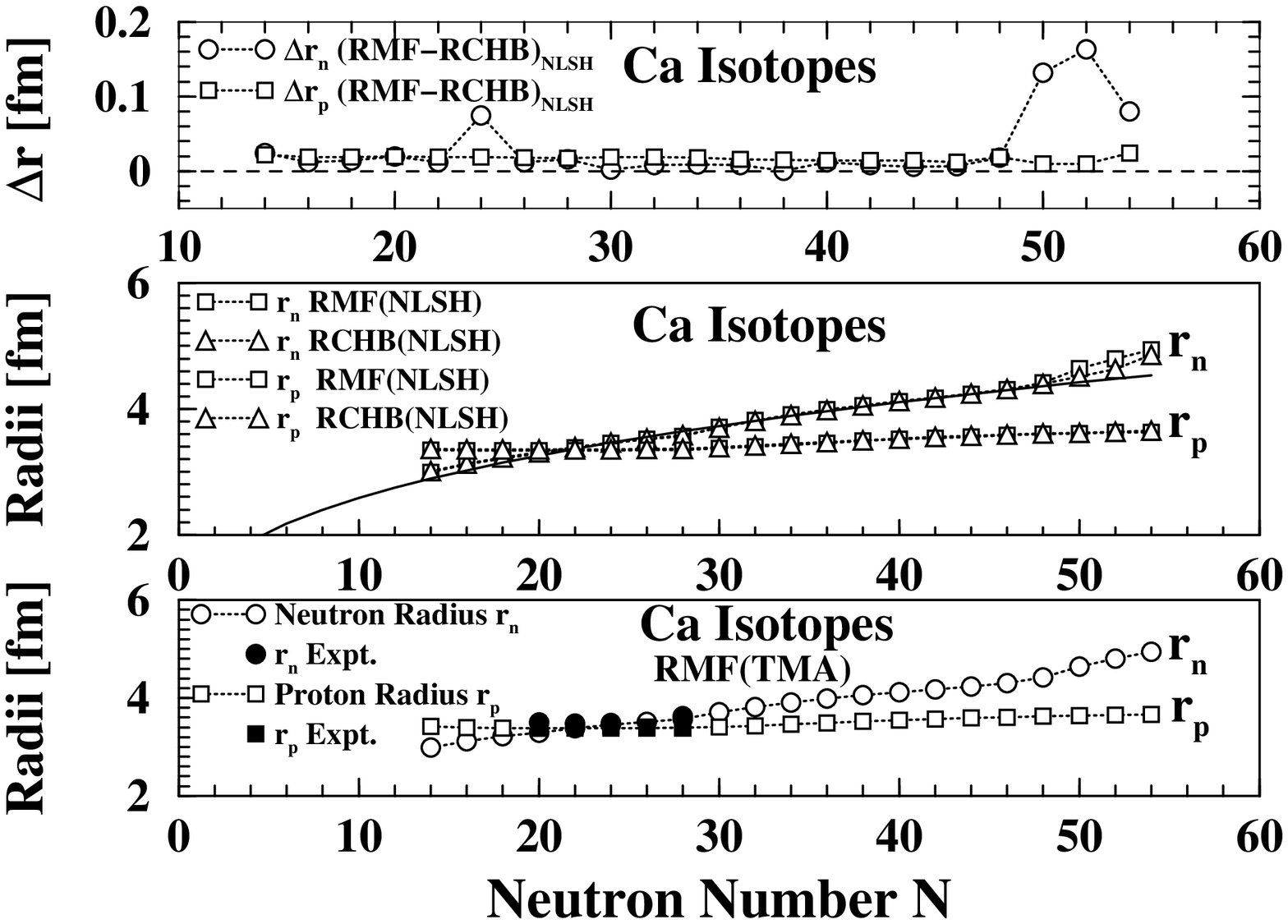,width=10.0
cm,height=8.0cm} \vskip 0.015in {\noindent \small {{\bf Fig.4.}
Lower panel:  The rms radii of neutron distribution $r_n$ (open
circles), and that of proton distribution $r_p$ (open squares)
obtained with the TMA force  are compared with the available
experimental data \cite{batty}, shown by solid circles and solid
squares, respectively. Middle panel: A comparison of RMF+BCS
results (open squares) for rms radii $r_n$ and $r_p$  with that of
RCHB (open triangles) from Ref.\cite{meng2} obtained with the
NL-SH force . Upper panel: Difference between the results obtained
from RMF+BCS and the RCHB approaches for the rms radii using the
NL-SH force shown in the middle panel. \mbox{} }}
\end{center}

As depicted in the upper panel of the Fig.6, the proton
distributions are observed to be confined to smaller distances.
Moreover, these start to fall off rapidly already at smaller
distances ( beyond $r> 3$ fm.) as compared to those for the
neutron density distributions shown in the lower panel. In the
interior as well as at outer distances, as shown in the inset of
the upper panel, the proton density values are larger for the
proton rich $\rm{Ca}$ isotopes and decrease with increasing
neutron number N. However, in the surface region, ($r \approx 4$
fm), the proton density values reverse their trend and increase
with increasing neutron number. Due to this feature of the proton
density distributions the proton radii are found to increase,
albeit in a very small measure, with increasing neutron number.

Similarly the lower panel of Fig.6, depicting the neutron density
distribution, shows that  for the magic numbers N = 14, 20, 28 and
40 the neutron densities fall off rapidly and have smaller tails
as compared to the isotopes with other neutron numbers. The
density distribution for the N = 50 case is seen to be very
different from the above cases indicating thereby that for the
neutron
\begin{center}
\psfig{figure=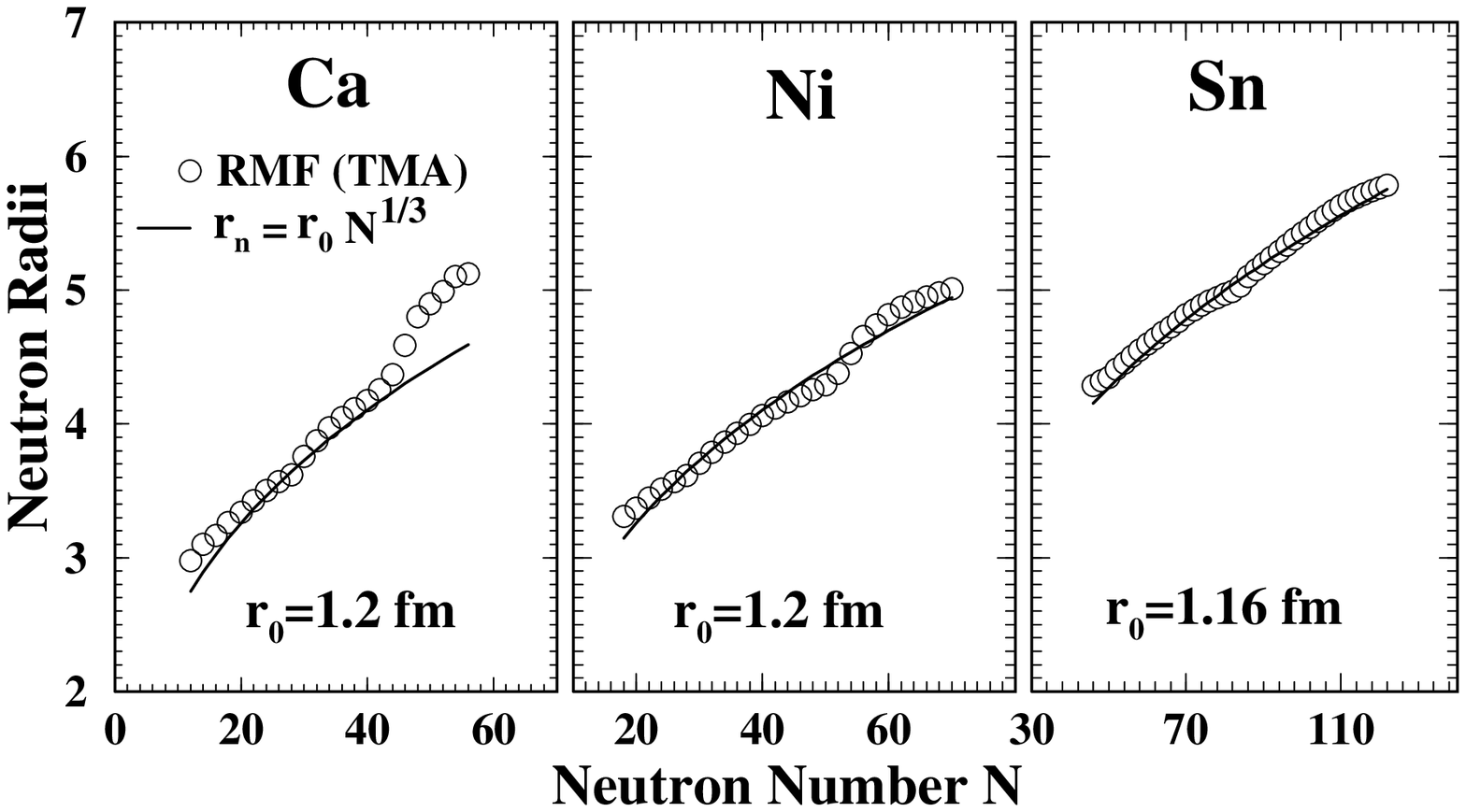,width=9.0 cm,height=7.0cm} \vskip
0.001in {\noindent \small {{\bf Fig.5.} The RMF
results\cite{yadav,yadav1,yadav2} for the neutron rms radii $r_n$
for the isotopes of $\rm {Ca}$, $\rm {Ni}$ and $\rm {Sn}$ nuclei
obtained with the TMA force. These are compared with a rough
estimate of neutron distribution radius given by $r_n$ = $r_0
N^{1/3}$ wherein the radius constant $r_0$ is chosen to provide
the best fit to the theoretical results. Halo formation in the
case of neutron rich $\rm {Ca}$ isotopes is clearly seen.\mbox{}
}}
\end{center}
rich $\rm {Ca}$ isotopes the N = 50 does not correspond to
a magic number. The neutron densities of the isotopes having $N >
40$ are found to exhibit especially widespread distributions out
side the range of the interaction potential. This has been
explicitly demonstrated in the inset of lower panel of Fig.6.
Moreover, these results are also found to be very similar to those
obtained using the RCHB approach\cite{meng2}. In particular, for
the isotopes with neutron shell closure corresponding to $N=$ 14,
20, 28 and 40 this similarity extends up to large radial
distances. For the other isotopes, there are small deviations
between the RMF+BCS and the RCHB approaches beyond the radial
distance $r = 8$ fm. However, beyond this distance the densities
are already quite reduced ranging between $10^{-4}$ fm$^{-3}$ to
$10^{-8}$ fm$^{-3}$. In Fig.6 (lower panel) it is interesting to
note that the neutron density distributions, out side the nuclear
surface and at large distances, for the neutron rich $\rm Ca$
isotopes with neutron number $N \ge 42$ are larger by several
orders of magnitude as compared to the lighter isotopes. This
behavior of the density distribution for the neutron rich $\rm Ca$
isotopes is quite different from the corresponding results,
especially  for the neutron rich isotopes of $\rm {Ni}$, $\rm
{Sn}$ and $\rm {Pb}$ nuclei\cite{yadav,yadav1,yadav2}. In the
latter cases, as the neutron number is added the tail of the
neutron density distributions for the neutron rich isotopes tend
to saturate.

The large relative enhancement in the tail region of neutron
density distributions of the neutron rich $\rm {Ca}$ isotopes
beyond $N = 40$ (or $A=60$) gives rise to the halo formation in
these isotopes. Essentially, it is caused due to weakly bound
neutrons occupying the single particle states near the Fermi level
which is itself almost close to zero energy for the neutron rich
isotopes as has been shown in Fig.7. The large energy gaps between
single particle levels $1d_{5/2}$ and $1d_{3/2}$ (not shown in
Fig. 7), and the levels $2p_{1/2}$ and $3s_{1/2}$  etc. are
responsible for the properties akin to shell or sub-shell closures
in the $^{34-72}\rm{Ca}$ isotopes for the neutron number N = 14
and 40 apart from the traditional magic nos. N = 20 and 28.
However the N = 50 shell closure is found to disappear due to
absence of gaps between the $1g_{9/2}$ state and the states in the
s-d shell. A better understanding of the halo formation in the
neutron rich $(N > 40)$ $\rm{Ca}$ isotopes is rendered if one
looks into the detailed features of single particle spectrum and
its variation shown in Fig.7 as one moves from the lighter isotope
to heavier one . For example, the neutron Fermi energy which lies
at $\epsilon_f = -19.90$ MeV in the neutron deficient
$^{34}\rm{Ca}$ nucleus moves to $\epsilon_f = -0.21$ MeV in  the
neutron rich $^{62}\rm{Ca}$, and to $\epsilon_f = 0.08$ MeV
(almost at the beginning of the single particle continuum) in
$^{70}\rm{Ca}$. The $1g_{9/2}$ state which lies at higher energy
in continuum for the lighter isotopes, comes down gradually to
become slightly bound for the neutron rich isotopes. Similarly,
the $3s_{1/2}$ state which lies in continuum for the lighter
isotopes (for example at $\epsilon = 0.70$ MeV in $^{34}\rm{Ca}$)
also comes down, though not so drastically, to become slightly
bound (($\epsilon = -0.05$ MeV in $^{68}\rm{Ca}$) for the neutron
rich isotopes.
\begin{center}
\psfig{figure=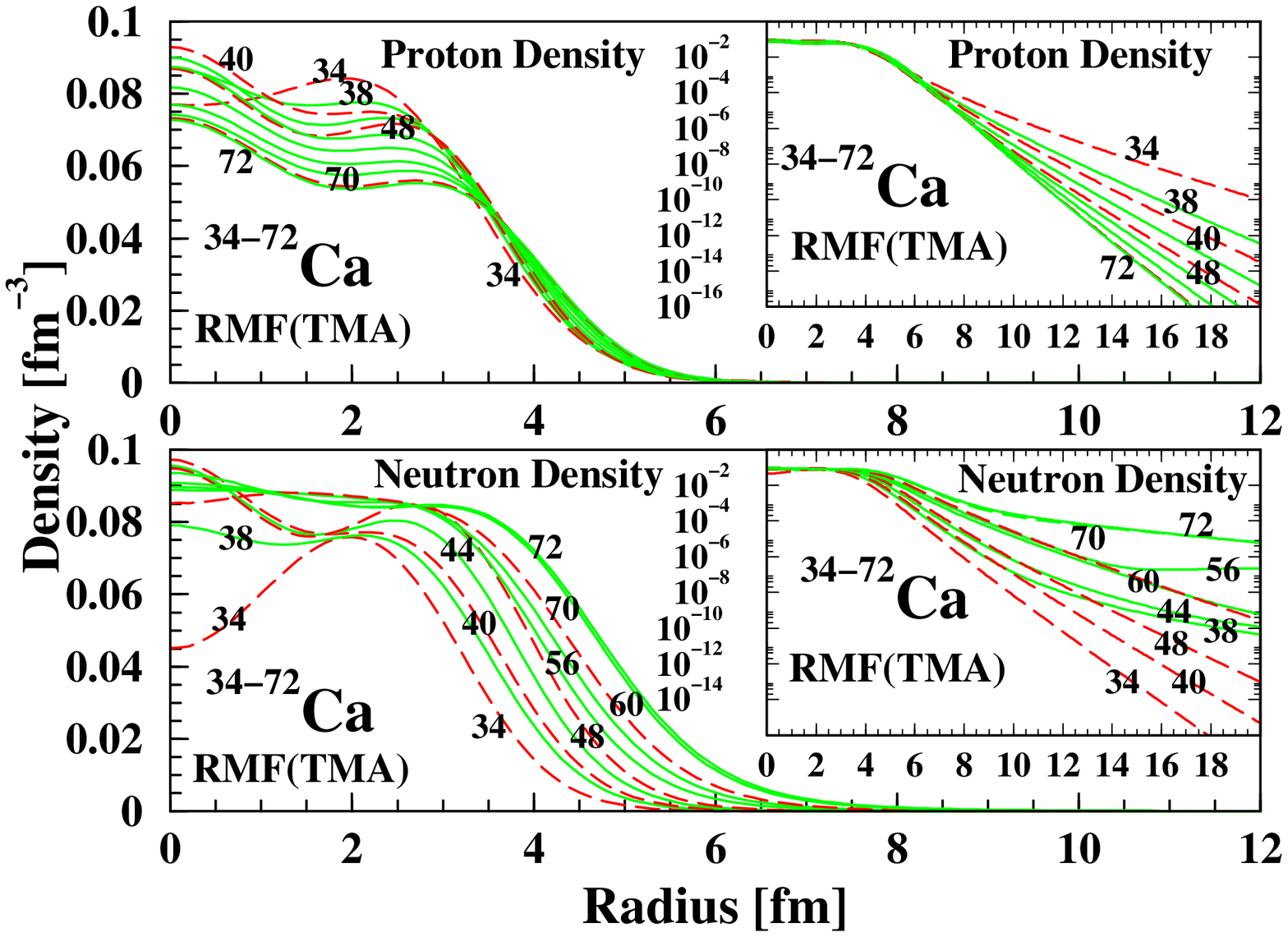,width=10.0cm,height=8.0cm}
\vskip 0.05in {\noindent \small {{\bf Fig.6.} The upper and lower
panels, respectively, show the proton and neutron density
distributions for the $\rm{Ca}$ isotopes obtained with the TMA
force. The numbers on the density distribution lines indicate the
mass number of the $\rm{Ca}$ isotope.The insets show the results
on a logarithmic scale up to rather large distances. \mbox{}}}
\vspace{0.05cm}
\end{center}

In the case of $^{60}\rm{Ca}$ with shell closure for both protons
and neutrons, the neutron single particle states are filled in up
to the $2p_{1/2}$ state, while the next high lying states
$3s_{1/2}$ and $1g_{9/2}$, separated by about $5$ MeV from the
$2p_{1/2}$ level, are completely empty. Now on further addition of
2 neutrons, it is observed that the $1g_{9/2}$ is filled in first
even though $3s_{1/2}$ state is slightly lower (by about 0.31 MeV)
than the $1g_{9/2}$ state as has been shown in the lower panel of
Fig.7. Still another addition of 2 neutrons is found to fill in
the $1g_{9/2}$ state once again, though  now the $1g_{9/2}$ state
is higher to the $3s_{1/2}$ state merely by $0.08$ MeV. This
preference for the $1g_{9/2}$ state stems from the fact that in
contrast to the $3s_{1/2}$ state, the positive energy $1g_{9/2}$
state being a resonant state has its wave function  entirely
confined inside the potential well akin to a bound state as shown
earlier in Fig.2 for the nucleus $^{64}\rm{Ca}$. For the neutron
number N = 48, it is found that both the $1g_{9/2}$ and $3s_{1/2}$
states become bound and start to compete together to get occupied
on further addition of neutrons as can be seen in Fig.7 (lower
panel). Further it is observed from the figure that both of these
states are completely filled in for the neutron number N = 52 and,
thus, the neutron drip line is reached with a loosely bound
$^{72}\rm{Ca}$ nucleus. The next single particle state,
$2d_{5/2}$, is higher in energy by about 0.5 MeV in the continuum
and further addition of neutrons does not produce a bound system.
\begin{center}
\psfig{figure=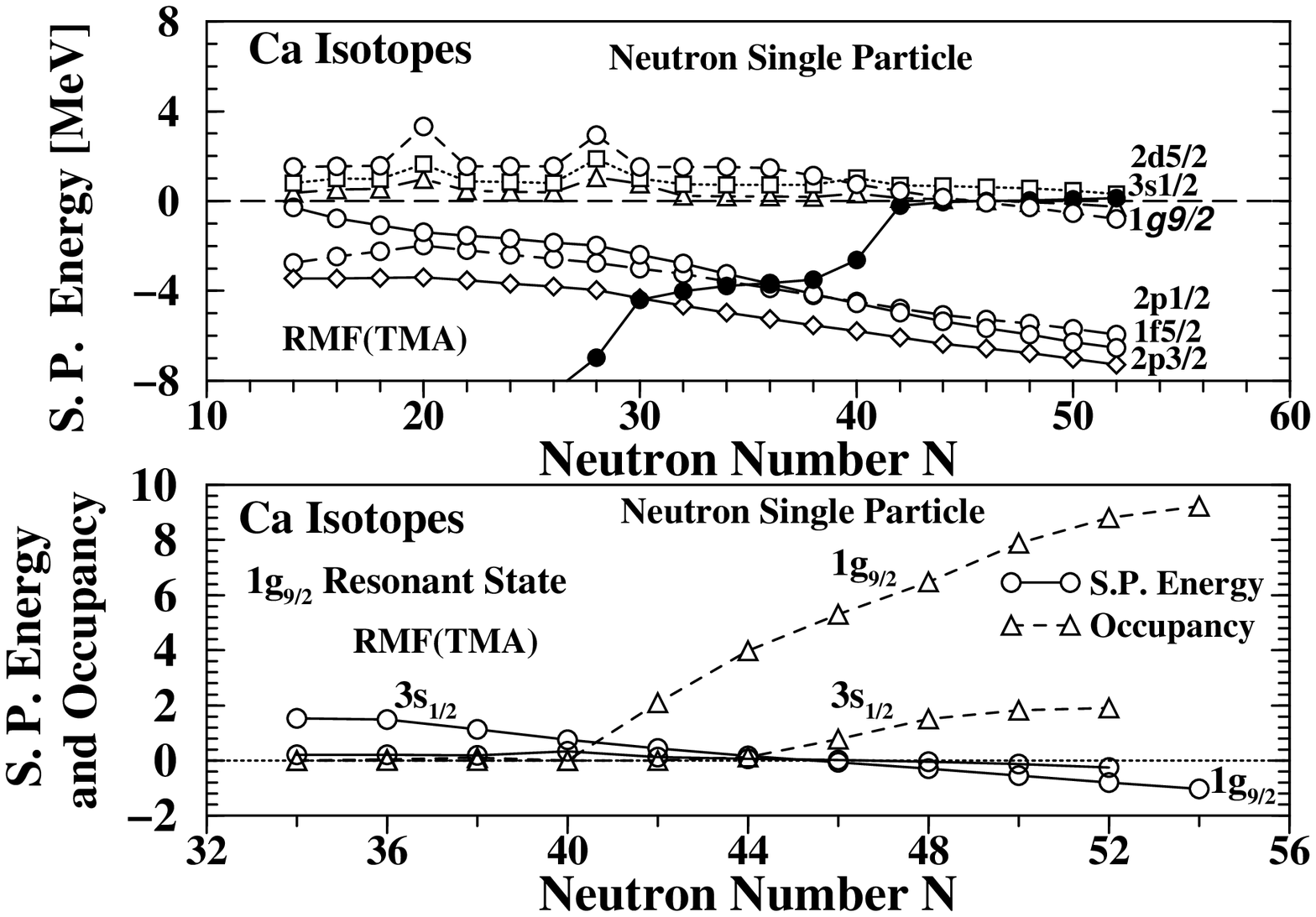,width=10.0cm,height=7.0cm}
\vskip 0.001in {\noindent \small {{\bf Fig.7.}Upper panel:
Variation of the neutron single particle energies obtained with
the TMA force for the $\rm{Ca}$ isotopes with increasing neutron
number. The Fermi level has been shown by filled circles connected
by solid line to guide the eyes. Lower panel: Variation of the
position and occupancy (no. of neutrons occupying the levels) of
the neutron $1g_{9/2}$ and $3s_{1/2}$ single particle states in
the neutron rich $\rm{Ca}$ isotopes.\mbox{} }} \vspace{0.001cm}
\end{center}

As mentioned above, the $1g_{9/2}$ state is mainly confined to the
potential region, and hence its contribution to the neutron radii
is similar to a bound state. In contrast, the $3s_{1/2}$ state
which has no centrifugal barrier and, therefore, is spread over
large spatial extension contributes substantially to the neutron
density distribution at large distances. Due to this reason, for
$N > 42$ as the $3s_{1/2}$ state starts being occupied the neutron
density distributions develop large tails, and the neutron radii
for the neutron rich isotopes ($N  > 42$) grow abruptly as has
been shown in figs. 4 and 5. Thus the filling in of the $3s_{1/2}$
single particle state with increasing neutron number in the
$^{64-72}\rm{Ca}$ isotopes causes the formation of neutron halos
in these nuclei.

It is pertinent to point out that our results described above are
consistent with the recent non-relativistic HFB calculations of
Bennaceur et al. \cite{dobac0} demonstrating the effects of
pairing correlations in the description of weakly bound neutron
rich nuclei. The interesting result of these authors \cite{dobac0}
is that the pairing correlations in the even-N nuclei, contrary to
the case of odd-N nuclei (N being the neutron number) for the zero
orbital angular momentum (l = 0) s-state, provide additional
binding in a way to halt the unlimited increase in the rms radius
as the single particle binding of the s-state tends to be zero.
Further, it has also been demonstrated by these authors
\cite{dobac0} that the low lying $l=0$ continuum states contribute
significantly in generating large rms neutron radii of the neutron
rich weakly bound nuclei, and that the exact position of l=0
orbital is not crucial for this enhancement.

From our Fig. 7 it is seen that the $3s_{1/2}$ orbital whose
positions varies only slightly for the neutron number $N=12$ to
$N=52$ continues to be close to zero energy, starts being occupied
only for $N > 42$. This occupancy generates large increase in the
neutron rms radii beyond $N=42$ as is readily observed in figs. 4
and 5 for the $Ca$ isotopes. Thus our results are in accord with
those of ref. \cite{dobac0} whereby the occupied $l=0$ orbital
provides a significant increase in the neutron rms radii and,
moreover, this increase does not tend to become infinitely large
even when the single particle energy of the s-state changes to lie
very close to zero. In a similar calculation for deformed nuclei
it may be difficult to discuss the results in terms of single
particle wave functions as shown in fig. 1 for the spherical case.
However, we believe that the main conclusions drawn here to
characterize a favorable situation for the halo formation will not
change significantly in a similar description for the even-even
deformed nuclei .

\section{Conclusion}

In conclusion, we have applied the BCS approach using a
descretized continuum within the framework of relativistic
mean-field theory to study the ground state properties of
$\rm{Ca}$ isotopes up to the drip-lines, the main emphasis being
on the possible formation of halos in the neutron rich $\rm{Ca}$
isotopes. Calculations have been performed using the popular TMA
and the NL-SH sets of parameters  for the effective mean-field
Lagrangian. For the pairing energy a $\delta$-function interaction
has been employed for the state dependent BCS calculations. It is
found that from amongst the positive energy states, apart from the
single particle states adjacent to the Fermi level, the dominant
contribution to the pairing correlations is provided by a few
states which correspond to low-lying resonances. An important
result to be emphasized is the following. In the vicinity of
neutron drip-line for the $\rm{Ca}$ isotopes, it is found that
further addition of neutrons causes a rapid increase in the
neutron rms radius with a very small increase in the binding
energy, indicating thereby the occurrence of halos in the neutron
rich $\rm{Ca}$ isotopes. The filling in of the resonant $1g_{9/2}$
state, that sets in  even before it becomes bound, with a very
little increase in the binding energy causes the existence of
extremely neutron rich $\rm{Ca}$ isotopes, whereas the occupancy
of loosely bound $3s_{1/2}$ state gives rise to the halo
formation. Also, as in earlier prototype
calculations\cite{yadav,yadav1} for the $\rm{Ni}$ and $\rm{Sn}$
isotopes, our present RMF+BCS results for the two neutron
separation energy, rms neutron and proton radii and pairing
energies for the $\rm{Ca}$ isotopes compare well with the known
experimental data\cite{audi}. Furthermore, detailed comparisons
show that the RMF+BCS approach provide results almost similar to
those obtained in the more complete relativistic continuum Hartree
Bogoliubov (RCHB) treatment\cite{meng2}. This is in accord with
the conclusion of Grasso et al.\cite{grasso} whereby the BCS
approach is shown to be a good approximation to the Bogoliubov
treatment in the context of non-relativistic mean-field studies.
Moreover, the effects of pairing correlations observed in our
treatment are found  to be in agreement with those demonstrated
recently by Bennaceur et al. \cite{dobac0} as regards to the
contribution of the $l=0$ orbital angular momentum s-state to the
large enhancement of the neutron rms radii, and also to the so
called pairing anti-halo effect which prevents the unlimited
growth of the neutron rms  radii when the single particle energy
of the $l=0$ orbital tends to be zero.

\subsection*{Acknowledgments} One of the authors (HLY)would like
to thank Prof. Toki for numerous fruitful discussions. Support
through a grant by the Department of Science and Technology(DST),
India is also acknowledged. HLY would also like to thank Prof.
Faessler for his kind hospitality while visiting Institut f\"{u}r
Theoretische Physik der Universit\"{a}t T\"{u}bingen, Germany
where part of this work was carried out. The authors are indebted
to J. Meng for communicating some of his RCHB results before
publication.



\begin{thebibliography}{99}
\bibitem{tanihata}  I. Tanihata,  J. Phys. G
{\bf 22} (1996) 157; I. Tanihata {\it et al}.,  Phys. Rev. Lett.
{\bf 55} (1985)  2676.
%
\bibitem{jensen}  A. S. Jensen, K. Riisager, Phys. Lett. B
{\bf 480} (2000) 39, and references therein.
%
\bibitem{ring0}  J. Meng,  P. Ring, Phys. Rev. Lett. B
{\bf 80} (1998) 460, and references therein.
%
\bibitem{dobac0}  K. Bennaceur, J. Dobaczewski and M. Ploszajczak,
Phys. Lett. B {\bf 496} (2000) 154, and references therein.
%
\bibitem{migdal73}  A. B. Migdal, Jour. Nucl. phys.
{\bf 16} (1973) 238.
%
\bibitem{dobac}  J. Dobaczewski, H. Flocard, and J. Treiner,
 Nucl. Phys. A {\bf 422}, \rm  103 (1984); J. Dobaczewski,
 W. Nazarewicz and T. R. Werner,
 Phys. Scr. T {\bf 56 }, \rm 15 (1995).
%
\bibitem{terasaki}  J. Terasaki, P.-H. Heenen, H. Flocard, and P. Bonche,
 Nucl. Phys. A  {\bf 600}, \rm 371 (1996).
%
\bibitem{dobac96}  J. Dobaczewski, W. Nazarewicz, T. R. Werner, J. F. Berger,
 C. R. Chinn, and J. Decharge,
 Phys. Rev. C {\bf 53}, \rm   2809 (1996).
%
\bibitem{sand}  N. Sandulescu, R. J. Liotta, and R. Wyss,
 Phys. Lett. B {\bf 394}, \rm 6 (1997).
%
\bibitem{sand2}  N. Sandulescu, Nguyen Van Giai and R. J. Liotta,
 Phys. Rev. C {\bf 61}, \rm 061301(R) (2000).
%
\bibitem{grasso}  M. Grasso, N. Sandulescu, Nguyen Van Giai
and R. J. Liotta, Phys. Rev. C {\bf 64}, \rm 064321 (2001).
%
\bibitem{walecka}  J. D. Walecka,  Ann. Phys. (N.Y.) {\bf 83},
\rm 491 (1974).
%
\bibitem{serot} B. D. Serot and J. D. Walecka,
 Adv. Nucl. Phys. {\bf 16}, \rm  1 (1986).
%
\bibitem{pgr} P.-G. Reinhard, M. Rufa, J. Marhun, W. Greiner and J. Friedrich,
 Z. Phys. A  {\bf 323},  \rm   13 (1986).
%
\bibitem{bouy}  A. Bouyssy, J.-F. Mathiot, Nguyen Van Giai, and S. Marcos,
 Phys. Rev. C {\bf 36}, \rm   380 (1987).
%
\bibitem{pgr2}  P-G Reinhard,  Rep. Prog. Phys.  {\bf 52},
\rm 439 (1989) and references therein.
%
\bibitem{gambhir}  Y. K. Gambhir, P. Ring, and A. Thimet,  Ann. Phys.
(N.Y.) {\bf 198}, \rm 132 (1990).
%
\bibitem{toki} H. Toki, Y Sugahara, D. Hirata,
B. V. Carlson, and I. Tanihata,
 Nucl. Phys. A {\bf 524}, \rm  633 (1991).
%
\bibitem{brock}  R. Brockman and  H. Toki,
 Phys. Rev. Lett.  {\bf 68}, \rm  3408 (1992).
%
\bibitem{hirata} D. Hirata, H. Toki,  I. Tanihata, and P. Ring,
 Phys. Lett. B {\bf 314},  \rm 168 (1993).
%
\bibitem{suga} Y Sugahara and H. Toki,
 Nucl. Phys. A {\bf 579}, \rm  557 (1994); Y Sugahara, Ph.D.
Thesis, Tokyo Metropolitan University, 1995.
%
\bibitem{ring}  P. Ring,   Prog. Part. Nucl. Phys.  {\bf 37},
\rm  193 (1996) and references therein.
%
\bibitem{sharma}  M. M. Sharma, M. A. Nagarajan and P. Ring,
 Phys. Lett. B {\bf 312}, \rm 377 (1993).
%
\bibitem{meng}  J. Meng,  Phys. Rev. C {\bf 57}, \rm  1229
    (1998).
%
%
\bibitem{meng4}  J. Meng,  Nucl. Phys. {\bf 635}, \rm
3 (1998).
%
\bibitem{lala}  G. A. Lalazissis, D. Vretenar and P. Ring,
 Phys. Rev. C {\bf 57}, \rm  2294 (1998).
%
\bibitem{mizu}  S. Mizutori, J. Dobaczewski, G. A. Lalazissis, W. Nazarewicz,
and P. G. Reinhard,  Phys. Rev. C {\bf 61}, \rm  044326
    (2000).
%
\bibitem{leja}  J. Leja, S. Gmuca,
 Acta Phys. Slov. {\bf 51}, \rm 201 (2001).
%
\bibitem{estal}  M. Del Estal, M. Contelles, X. Vinas, and S. K. Patra,
 Phys. Rev. C {\bf 63}, \rm  044321 (2001).
%
\bibitem{yadav}  H. L. Yadav, S. Sugimoto and H. Toki,
 Mod. Phys. Lett. A {\bf 17}, \rm 2523 (2002);\ Preprint, RCNP,
Osaka University, Osaka (2001).
%
\bibitem{yadav1}  H. L. Yadav, U. R. Jakhar and K. C. Agarwal,
 Acta Phys. Slov. {\bf 53}, \rm 25 (2003).
%
\bibitem{meng2}  J. Meng, H. Toki, J.Y. Zeng, S. Q. Zhang and S. Q. Zhou,
 Phys. Rev. C {\bf 65}, \rm 041302(R) (2002).
%
\bibitem{yadav2}  H. L. Yadav, M. Kaushik and H. Toki,
Int. J. Mod. Phys. {\bf E 13},\rm 647 (2004)
%
\bibitem{lane}  A. M Lane,  Nuclear Theory (Benjamin, 1964).
%
\bibitem{ring2}  P. Ring and P. Schuck,
 The Nuclear many-body Problem (Springer, 1980).
%
\bibitem{bertsch91}  G. F. Bertsch and  H. Esbensen,
 Ann. Phys. (N.Y.)  {\bf 209} (1991) 327.
\bibitem{migdal67}  A. B. Migdal, Theory of Finite Fermi Systems
and Applications to Atomic Nuclei (Interscience, New York, 1967).
%
\bibitem{audi}  G. Audi, A. H. Wapstra and C. Thibault,
 Nucl. Phys. A {\bf 729}, \rm 337 (2003).
%
\bibitem{vries}  H. de Vries, C. W. de Jager, and C. de Vries,
 At. Data Nucl. Data Tables {\bf 36}, \rm  495 (1987).
%
\bibitem{batty}  C. J. Batty, E. Friedman, H. J. Gils, and H. Rebel,
 Adv. Nucl. Phys. {\bf 19}, \rm 1 (1989).
%
\end{thebibliography}
\end{document}